\documentclass{vospwks2007}
\usepackage{graphicx}

\title{stellar synthetic spectroscopy in the Virtual Observatory era}
\author{Bertrand Plez}
\affil{GRAAL, CNRS, Universit\'e Montpellier 2, France}

\begin{document}


\maketitle

\begin{abstract}
Arguments for including in the VO large grids of 
synthetic spectra fully covering the HR diagram are presented.
One obvious need is that of population synthesis at all redshifts.
Theoretical spectra have also the power to predict peculiar behaviors,
that could remain unnoticed in observed spectra, or lead to erroneous 
conclusions. One interesting example is given of the Ca II H \& K lines in  
extremely metal-poor stars. In carbon-rich atmospheres, the H \& K
lines become much weaker relative to the continuum, which will
lead to an underestimate of the metallicity. What actually happens
is a displacement of the continuum, obvious in absolute-flux synthetic spectra, but not visible in observed 
continuum-normalized spectra. 
Libraries of synthetic spectra are 
described, as well as the codes used to compute them. A few remarks are 
made on issues with making these libraries available through the VO, 
as well as on the necessary input data, i.e. line lists and model
atmospheres. 
\end{abstract}

\section{Why synthetic spectroscopy in the VO ?}
Synthetic spectra have several advantages over observed spectra:
they naturally provide absolute fluxes (and intensities!), 
as well as normalization to the continuum, and 
they can be computed for whatever stellar parameters (even rare or unobserved 
types!). In principle, a database of synthetic spectra can cover 
the whole HR diagram at all metallicities, and non-standard chemical compositions,
which is invaluable for stellar population synthesis. Synthetic spectra may be used  
for the preparation of instruments, e.g. GAIA, or to understand characteristics 
of sub-samples in large surveys (using e.g. color-color diagrams
or low-resolution spectra). They are necessary to quickly 
classify spectra in large surveys (e.g.
micro-lensing searches) with non-standard wavelength ranges at 
various spectral resolutions.
And of course synthetic  spectra are necessary to extract stellar parameters from observations ($T_{\rm eff}$, [Fe/H], etc).

Various kinds of spectra can be computed, corresponding to the usage that is intended:
population synthesis may require low resolution, full spectral coverage, or conversely
a medium resolution, shorter spectral coverage (e.g. H band). High resolution, over a short
spectral interval is required to prepare the GAIA RVS (Ca II IR triplet region). Exploitation of VLT-UVES spectra requires very high resolution over the whole optical  and near-IR domain. In all cases, the variation of many parameters is a key to the superiority
of synthetic spectra over observations ($T_{\rm eff}$, gravity, 
[Fe/H], [$\alpha$/Fe], C, N, O abundances, etc). This may 
represent a large number 
of spectra if a significant portion of the HR diagram is to be covered 
(at least $10^3$ to $10^4$ spectra). The computation time of such an
amount of spectra is not a 
big issue if the model atmospheres used are classical (in the sense 1-D, LTE, 
hydrostatic).
If one at least of these simplifying hypotheses is lifted the computing time
may become prohibitively large. For some regions of the HR diagram, or if a great 
precision is wanted, this proves necessary (e.g NLTE and wind hydrodynamics for OB stars,
 non-LTE and 3-D hydrodynamical models of convection for better than 5\% accuracy in 
abundances in solar type stars). What is always necessary is a good set of physical 
input data (line lists, partition functions, etc).

\section{What can be done with synthetic spectra: four illustrations}
\subsection{impact of stellar parameters on spectra}
\begin{figure}
\centering
\includegraphics[width=0.8\linewidth,angle=-90]{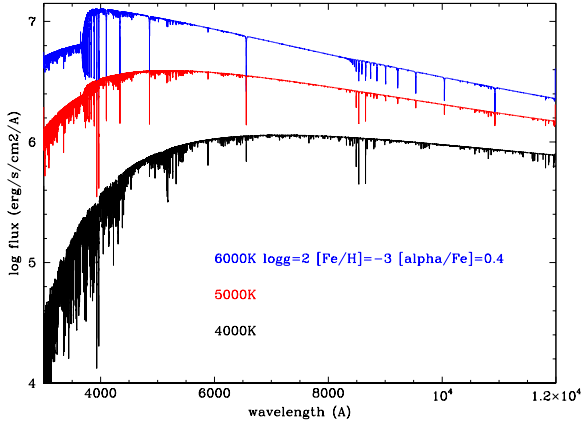}
\caption{Effect of $T_{\rm eff}$ variations on the optical and NIR spectrum
of extremely metal-poor giant model stars (MARCS models). Note the disappearance
of Hydrogen lines (and jumps) and the strenghtening of molecular lines (e.g. CH around 4300\AA\ with decreasing
temperature. All spectra on an absolute flux scale at the stellar surface.\label{fig-teff}}
\end{figure}
\begin{figure}
\centering
\includegraphics[width=0.8\linewidth,angle=-90]{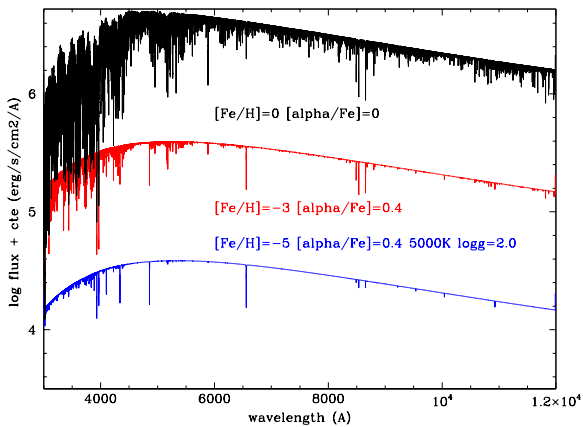}
%\caption{Effect of [Fe/H] variations on the optical and NIR spectrum
%of extremely metal-poor giant model stars (MARCS models). 
\caption{Same as Fig.~\ref{fig-teff} for variations of [Fe/H]. Note the weakening
of the Ca II triplet around 8500{\AA}.\label{fig-met}}
\end{figure}
\begin{figure}
\centering
\includegraphics[width=0.8\linewidth,angle=-90]{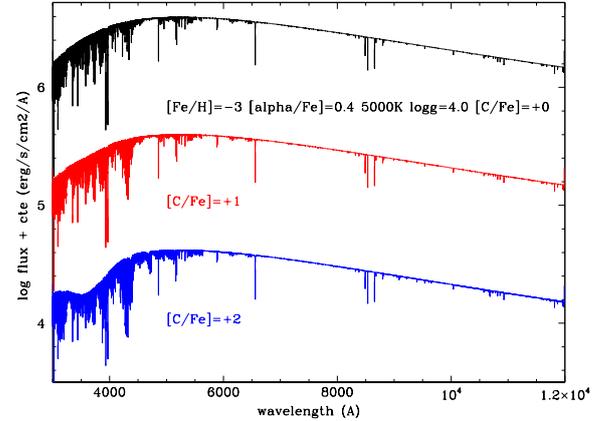}
%\caption{Effect of carbon abundance variations on the optical and NIR spectrum
%of extremely metal-poor giant model stars (MARCS models). 
\caption{ Same as Fig.~\ref{fig-teff} for carbon abundance variations. Molecular lines become
progressively prominent at larger [C/Fe] ratios: CH around 4300\AA, or C$_2$ at larger wavelengths. Note also the distortion of the UV continuum.\label{fig-cabund}}
\end{figure}

I chose here to illustrate this on some unusual extremely metal-poor  carbon-rich stars. Carbon-enhanced stars make up around 20\% of very metal-poor stars
and have become a boiling field of research in recent years (see the review of 
\citet{beers:2005}). The model atmospheres used for the calculation of 
the spectra shown on Fig.~\ref{fig-teff}, ~\ref{fig-met}, and ~\ref{fig-cabund} were
computed with the MARCS code \citep{gust:1975, plez:1992, gust:2003}. The Figures speak for themselves. Similar observed spectra are used to
search for metal-poor
stars with specific characteristics of temperature, metallicity or carbon 
enhancement in low-resolution spectroscopic surveys like the SDSS-SEGUE (Sloan 
Extension for Galactic Understanding and Exploration, see e.g.  
\citet{siva:2005}). Also, synthetic photometry can be directly computed from 
these spectra and used for similar purposes. One particular interest is the 
calibration of color-temperature relations for extremely metal-poor stars,
especially of non-standard composition, that are only observed in small 
 numbers, if at all.
 
\subsection{Effect of carbon enhancement on the Ca II H \& K lines}
\begin{figure}
\centering
\includegraphics[width=0.8\linewidth,angle=-90]{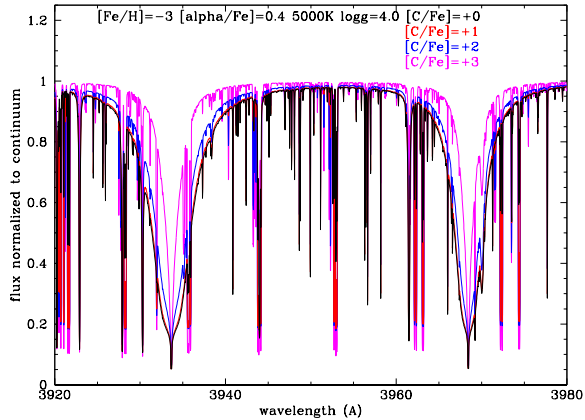}
\caption{Effect of carbon abundance variations on the Ca II H \& K lines in
an extremely metal-poor giant model star (MARCS models). The Ca 
abundance is the same in all models. Flux normalized to the continuum,
as would be the case for observed spectra.\label{fig-czoomrel}}
\end{figure}

\begin{figure}
\centering
\includegraphics[width=0.8\linewidth,angle=-90]{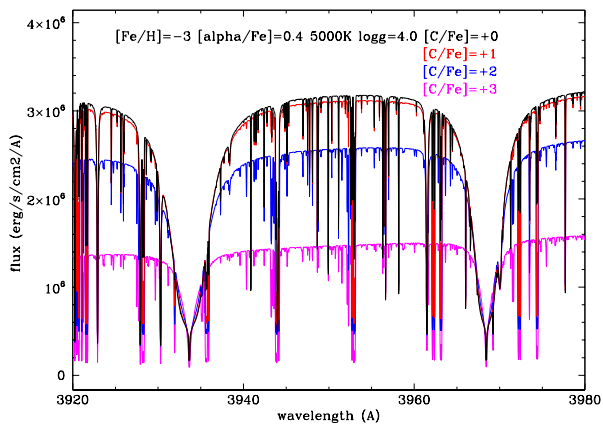}
\caption{Same as fig.~\ref{fig-czoomrel}, but on an absolute flux
scale. Note that the continuum level is very different in the four
models, but the Ca II line cores are unaffected. The Ca abundance
is the same, only the equivalent width of the lines is much smaller
in the C-rich spectra.\label{fig-czoomabs}}
\end{figure}

Fig.~\ref{fig-czoomrel} shows the near-UV model spectra of extremely 
metal-poor
main sequence stars with  $T_{\rm eff}$=5000K. The Ca II lines appear 
weaker with higher C/Fe ratio, which gives the impression at first glance
that the Ca abundance is lower. As many searches for metal-poor stars have
 relied  upon H \& K strength (in particular the many candidates extracted from
the HK survey of \citet{beers:1985}), it is important to understand 
this H \& K weakening to be able to disentangle Ca and C abundance effects.
What happens is understood by looking at the opacity in the continuum and line 
forming region in the models.
In the [C/Fe]=0 model atmosphere, the dominant continuum opacity source around 3500\AA\
is H$^-$, whereas in the model with [C/Fe]=2, the CH continuous absorption increases the 
opacity by a factor 10 to 100. This 
strong difference in the continuous opacity deeply affects the continuum 
flux, whereas line cores, with much larger opacities, remain mostly unchanged.
This can be seen in Fig.~\ref{fig-czoomabs} (absolute flux showing the change in continuum level),
as well as in  Fig.~\ref{fig-cabund} (distortion of the continuum in the 
near-UV). The effect is only seen because the flux is plotted on an
absolute scale, allowing comparison between spectra of stars with different
abundances. When observing at the telescope, the situation would look as in
 Fig.~\ref{fig-czoomrel}, and therefore look as a Ca underabundance, not a C
overabundance! Only synthetic spectra can reveal such an effect.
\subsection{Other illustrations}
Further illustration is given by the possibility to predict theoretical 
equivalent widths of, e.g., Balmer lines as a function of $T_{\rm eff}$ 
in hot stars,
useful in population synthesis, as shown by \citet{martins:2005}.
A last example is given by the work of \citet{massey:2005} and \citet{levesque:2005, levesque:2006},
using synthetic MARCS spectra and observed spectro-photometry of red supergiant
stars to derive simultaneously the $T_{\rm eff}$ of the stars and the
reddening on the line of sight. The reddening law affecting the slope
of the spectrum, does not change the strength of the TiO bands relative to
the local continuum. As this strength is very sensitive to temperature, 
the $T_{\rm eff}$ can be derived independently of the reddening, that 
follows from an adjustment of the slope of the theoretical spectrum on
the observed spectro-photometry. This work has lead to a revision of the 
temperature scale of red supergiants. Another result is that the 
reddening towards red supergiants is larger than in the surrounding
OB associations, which shows that there is a circumstellar dust contribution
to the reddening. The slope of the reddening indicates that the dust
grains are of larger size than their interstellar counterparts. Such a 
work would not be possible without synthetic spectra, as virtually all
red supergiants are reddened by unknown amounts, presumably by grains of various sizes.
\section{Some challenges for synthetic spectroscopy in the VO}
Some of the challenges ahead are, .e.g., (i) the analysis of huge amounts 
of spectra from 
GAIA, SDSS-SEGUE, RAVE, and multifibre spectrographs, esp. from ELTs, each of
which represents millions to billions of spectra, or (ii) to synthesize
 stellar populations at high redshifts, low metallicity, in various
  wavelength regions,
For that, we need very large grids of synthetic spectra, and fast,
 reliable and accurate algorithms to do it automatically. Two approaches may be chosen: to use large libraries of spectra from the VO to compare to
 observations with VO tools or to compute specific spectra for each 
 observation , either locally with resident software and computer, but using
 data and other tools from the VO, or with codes running on a server 
 inside the VO. We will further comment on these two philosophies 
 below.
\section{Libraries of synthetic spectra}
\begin{figure}
\centering
\includegraphics[width=0.8\linewidth,angle=-90]{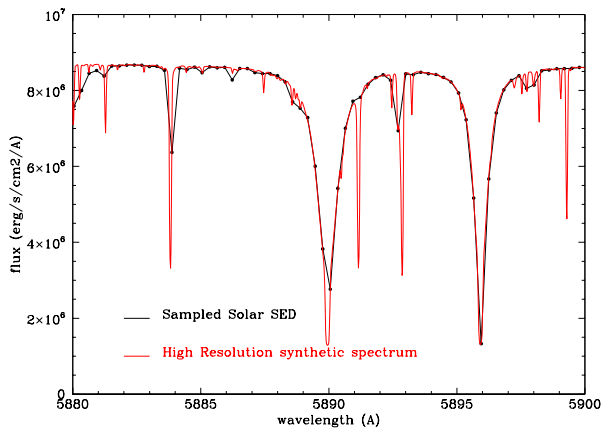}
\caption{Sampled SED from a MARCS model compared to a high
resolution spectrum computed with the same model and almost 
the same line data. The flux at each sampled wavelength is 
the same as  that computed at high resolution, but it gives no
clue to how the flux varies between points. A local integration
of an equivalent width will be erroneous, but on large scale
the average flux 
level is correct. \label{fig-sed}}
\end{figure}

A number of libraries already exist on the web, although not always 
integrated in the VO. Spectral Energy Distributions (SED) are 
available for the various brands of 1-D, static, LTE models ATLAS 
\citep{kurucz, castelli}, 
MARCS (marcs.astro.uu.se), and PHOENIX \citep{hauschildt:1996, 
hauschildt:1999}. Note that in some cases the SEDs are actually sampled
spectra, as the models are usually computed with the opacity
sampling algorithm, a Monte-Carlo approach to the radiative field
calculation. All radiative quantities are computed at
a large number of wavelengths, but nothing is known in between. 
The difference between a high resolution spectrum, and a 
spectrum sampled at a resolution of $R=\lambda/\Delta\lambda=20000$ is
shown in Fig~\ref{fig-sed}. It is clear that the sampled
flux cannot be compared to an observed spectrum at the same
$R=20000$ resolution. This can be done only at a degraded resolution.
A sampled spectrum will globally provide the right fluxes at a much 
lower resolution, but not locally at high resolution.
Detailed high-resolution spectra are only available for subsets of the
existing grids. There are too many to cite them all, but most can be 
found (together with observed libraries) on the excellent site of
D. M. Guti\'errez (www.ucm.es/info/Astrof/invest/actividad/spectra.html).
A problem still to be solved for many of these libraries is to 
properly include them in the VO (see one approach in the contribution 
by \citet{rodrigo}
in this volume). One library of synthetic spectra that is just about to 
be made available is the POLLUX database (see the contribution by 
\citet{pollux} in this volume). It will be directly VO compliant and
includes high resolution synthetic spectra (300-1200~nm),
covering most of the HR diagram, at all metallicities. It will also
provide SEDs at lower resolution. Regular updates are planned.
\subsection{Issues to keep in mind}
Computing such large grids of models and spectra may become 
time-consuming. CPU time is of course needed, but also scientist 
time to assemble 
the input data, set up the calculations, and check that the output
makes sense. Making the data available on the web, other than as a 
tar-file on an anonymous ftp, and especially if it has to be VO compliant
may also become a burden. Here, help from our friends of the VO is
critical. The benefits of getting ones models easily accessible 
on the web, are twofold: (i) the models get used and one hopefully 
gets acknowledged (the concern is that one has to find a way that
when data is made available through the VO, and is passed from 
one tool to another to finally become part of some end product, proper 
referencing to the original source is ensured), (ii) as importantly,
the use of the models towards different purposes, sometimes not anticipated by their producers, leads to a thorough tests of these models
and their limits. However, users must remember that model spectra may
be flawed, because of inappropriate physics (e.g. LTE O star atmospheres, 
1D static models for convective 3D atmospheres), because of inaccurate 
input data (e.g. line positions and strengths), or because of missing data 
(e.g. cool stars without TiO). It is important to understand the limits 
of a given brand of models and to refrain from using them to far off limits.
Differences will always appear at some level between observations
and calculations, but the problems are often not easy to ascribe 
to either models or observations. Observers tend to blame the models,
and vice-versa... One way to estimate systematic errors in models 
may be to compare models computed with different codes and input data
sets. 
While not always possible this is certainly very instructive. 
Illustrations  can be found in \citet{martins:2005}. 
Differences tend to increase for more extreme spectral types, and progress
needs to be made particularly on carbon stars, and very cool stars, 
although successes like the analysis of very cool AGB star spectra 
with $T_{\rm eff}$ down to 2700~K by \citet{anibal} demonstrate 
the level of sophistication of modern model atmospheres and spectra.
\section{Synthetic spectroscopy at home or Spectra on Demand from VO ?}
To address the challenges outlined above, one approach would be to 
implement in the VO interfaces to synthetic spectroscopy codes that
could be run on demand  by the user for specific combinations of 
stellar parameters. This would be very flexible, but prohibitively
demanding on resources
due to the huge amount of computations that could be required (new 
calculation for each slightly different stellar parameter or abundance).
A better way of doing is to provide in a simple and standard way
large grids of pre-computed spectra, at very high resolution (it is easy
to degrade them to lower resolutions), with only the main stellar parameters varied ($T_{\rm eff}$, gravity, Fe/H, $\alpha$-elements/Fe,
and possibly some CNO variations). These grids can be interpolated to 
produce population syntheses. All present needs in population synthesis
can presumably be fulfilled with these basic grids. 
Similarly, to analyse large numbers 
of observed spectra, the calculated grids can be interpolated
to feed automatic analysis codes, resulting in first estimates of the
main stellar parameters. Refinements can be made locally using a
spectrum synthesis code, model atmospheres, and spectral line lists.
The automatic tools for stellar parameters determination are certainly 
something one would want to see integrated into the VO. The development
of these codes has followed various lines: minimum distance methods,
that are limited to few parameters, neural networks that may be difficult
to teach, and also new solutions that seem fast and efficient like
MATISSE \citep{recio}.
\subsection{Synthetic spectroscopy codes (LTE)}
If the user wants to compute  spectra with specific stellar
parameters, we have seen that the only viable solution is a local
implementation of 
a synthetic spectroscopy code. However, ingredients to compute 
synthetic spectra may be found on the web, in the VO: stellar model
atmospheres, and atomic and molecular line lists (see below).
A number of synthetic spectroscopy codes are available from their 
authors. I will here only present the 1-D LTE codes that are most widely used
(multi-D codes, or polarized radiation codes are usually not freely
available; some non-LTE codes can be found on the web).
SYNTHE was developed by R. Kurucz and is adapted to the calculation
of spectra for the ATLAS models. It is available on line either on 
Kurucz page (kurucz.harvard.edu), or better, with updates and 
documentation, on wwwuser.oat.ts.astro.it/atmos. MOOG was initially 
written by C. Sneden, who continues to update it. The present version
is pure-LTE (no continuum scattering included), and intended
for FGK stars. It is available 
with documentation  on verdi.as.utexas.edu/moog.html. SME (spectroscopy 
made easy) is developed by N. Piskunov, and is further described in
his contribution in this volume. It allows in particular abundance 
stratifications. It is available upon request. TURBOSPECTRUM was developed
by B. Plez from an older code of the Uppsala group (B. Gustafsson). The
assumptions are similar to the MARCS code and the input physics is
the same. It is optimized for the computation of cool star spectra
(FGKMSC) and can compute large chunks of spectra with millions of 
spectral lines. It is available upon request, without documentation.
Some issues have to be kept in mind when using off-the-shelf codes.
Making a code available to a wide audience is not a simple task, if one
wants to avoid its use in conditions it was not intended for. Many
warnings and conditional tests have to be implemented (this is
usually not done, and the result will be that the code may run and produce 
an erroneous output without any warning). Writing a documentation is 
a tedious task. It may be incomplete, and is often not read thoroughly. 
Ideally, there should be continued contact between users and providers 
of codes in order to help improve, correct, debug and document them.

\subsection{Spectral line databases}
A main ingredient for the computation of synthetic spectra is the line
lists. There are many databases on the web. There are also lists of
databases, always incomplete, but useful: www.cfa.harvard.edu/amdata/ampdata/amdata.shtml. The Kurucz home page is an enormous resource, 
but not always well documented with regards to, e.g., what input data
was used to construct the lists. An excellent atomic line database 
is the VALD \citep{vald1, vald2, pisk}. A third release is planned for the Fall
2007, including for the first time molecular lines (TiO). 
Molecular line lists are scattered 
on various sites: NASA-Ames, University College London, ... 
A good source of information is the stellar atmosphere code
descriptions, as their conceptors strive to use the best and most 
complete data. 
The contribution by E. Roueff in this volume provides also a few more  
references.
\section{Summary and a few questions}
I have shown that the setup of large libraries of synthetic spectra
is needed in the VO and quite straightforward to implement 
if only a few stellar parameters are allowed to vary. Of course
the data model needs to be finalized, but the discussions we had at
the conference will 
certainly speed up the process. For example, calculated spectra cannot
be queried with a position on sky, which is the usual way to 
query observations... On the other hand one would like to be able
to query synthetic spectra with an abundance pattern, or a temperature.
Such libraries will allow to address many challenges of modern astronomy, but
more detailed investigations (e.g. with specific chemical composition) 
based on synthetic spectra require a local
implementation of synthetic spectroscopy codes, that are available  
off-the-shelf. The input data (line lists) 
is not yet  consistently
included in the VO, and the data model is still under development. 
The situation is better for atomic data, with the existence of the VALD 
database. Molecular data is scattered on the web and the best way
so far to assemble line lists is to contact people that have done
it for their own work. 
Some challenges are ahead, like the automatic classification of spectra,
and the determination of stellar parameters on samples of billions 
of spectra. New methods are being tested that may allow to do this 
reliably on a large scale. These tools could be included in the VO. 
Other tools that should be included in the VO are correlation and
convolution (with rotation, macroturbulence, and instrumental profiles)
tools that could be used on both observed and 
calculated spectra. And finally, n all instances one has to make sure
that proper credit is given to all developers of such tools, models and
calculations, if one wants to keep alive a spirit of sharing resources
in the community. Credit is unfortunately needed to get jobs and money.
An possible solution was given at the conference: forward the information
on developers in the metadata sent to users, and inform journal editors and referees 
that they should ensure proper referencing of the source of data 
extracted from the VO. And finally, it is not easy to go mining into 
calculated (or observed data), and we will have to keep on learning to 
use tremendous amounts of data in an educated way.
\section*{Acknowledgments}
I thank the organizers for inviting me to this meeting,
allowing me to get in touch with the VO world, understand some of it,
and hopefully transmit some useful information and constraints 
for the inclusion of synthetic spectra into the VO.

\end{document}